# DISTRIBUTED VIDEO CODING: CODEC ARCHITECTURE AND IMPLEMENTATION

Vijay Kumar Kodavalla[1] and Dr. P.G. Krishna Mohan[2]

[1] Semiconductor and Systems Division, Wipro Technologies, Bangalore, India
vijay.kodavalla@wipro.com

[2] Electronics & Communications Engineering Dept, JNTU College of Engineering, Hyderabad, India
pgkmohan@yahoo.com

## ABSTRACT

*Distributed Video Coding (DVC) is a new coding paradigm for video compression, based on Slepian-Wolf (lossless coding) and Wyner-Ziv (lossy coding) information theoretic results. DVC is useful for emerging applications such as wireless video cameras, wireless low-power surveillance networks and disposable video cameras for medical applications etc. The primary objective of DVC is low-complexity video encoding, where bulk of computation is shifted to the decoder, as opposed to low-complexity decoder in conventional video compression standards such as H.264 and MPEG etc. There are couple of early architectures and implementations of DVC from Stanford University[2][3] in 2002, Berkeley University PRISM (Power-efficient, Robust, hIgh-compression, Syndrome-based Multimedia coding)[4][5] in 2002 and European project DISCOVER (DIStributed COding for Video SERvices)[6] in 2007. Primarily there are two types of DVC techniques namely pixel domain and transform domain based. Transform domain design will have better rate-distortion (RD) performance as it exploits spatial correlation between neighbouring samples and compacts the block energy into as few transform coefficients as possible (aka energy compaction). In this paper, architecture, implementation details and "C" model results of our transform domain DVC are presented.*

## 1. INTRODUCTION

The encoding process of DVC is very simple by design. First, consecutive frames in the incoming video sequence are split into various groups based on the cumulative motion of all the frames in each group crossing a pre-defined threshold. And the number of frames in each such group is called a GOP (Group of Pictures). In a GOP, first frame is called key frame and remaining frames are called Wyner-Ziv (WZ) frames. The so called key frames will be encoded by using H.264 main profile intra coder. Higher GOP size increases the number of WZ frames between key frames, which reduces the data rate. The so called WZ frames will undergo block based transform, i.e., DCT is applied on each 4x4 block. The DCT coefficients of entire WZ frame will be grouped together, forming so called DCT coefficient bands. After the transform coding, each coefficient band will be uniform scalar quantized with pre-defined levels. Bit-plane ordering will be performed on the quantized bins. Each ordered bit-plane will be encoded separately by using Low Density Parity Check Accumulator (LDPCA) encoder. LDPCA encoder computes a set of parity bits representing the accumulated syndrome of the encoded bit planes. An 8-bit Cyclic Redundancy Check (CRC) sum will also be sent to decoder for each bit plane to ensure correct decoding operation. The parity bits will be stored in a buffer in the encoder and progressively transmitted to the decoder, which iteratively requests more bits during the decoding operation through the feedback channel.

The DVC decoding process is relatively more complex. The key frames will be decoded by using H.264 main profile intra decoder. The decoded key frames will be used for reconstruction of so called side information (SI) at the decoder, which is an estimation of the WZ frame available only at the encoder. A motion compensated interpolation between the two closest reference frames is performed, for SI generation. The difference between the WZ frame and the corresponding SI can be correlation noise in virtual channel. An online Laplacian model is used to obtain a good approximation of the residual (WZ-SI). The DCT transform will be applied on the SI and an estimate of the coefficients of the WZ frame are thus obtained. From these DCT coefficients, soft input values for the information bits will be computed, by taking into account the statistical modeling of the virtual noise. The conditional probability obtained for each DCT coefficient will be converted into conditional bit probabilities by considering the previously decoded bit planes and the value of side information. These soft input values will be fed to the LDPCA decoder which performs the decoding operation. The decoder success or failure will be verified by an 8 bit CRC sum received from encoder for current bit plane. If the decoding fails, i.e., if the received parity bits are not sufficient to guarantee successful decoding with a low bit error rate, more parity bits will be requested using the feedback channel. This is iterated until successful decoding is obtained. After the successful decoding of all bit planes, inverse bit plane ordering will be performed. Inverse quantization and reconstruction will be performed on the decoded bins. Next, inverse DCT (IDCT) will be performed and each WZ frame is restored to pixel domain. Finally, decoded WZ frames and key frames are interleaved as per GOP to get the decoded video sequence.

The previous work in this field includes pixel domain DVC architecture from Stanford University[2][3] in 2002 followed by Syndrome based transform domain based DVC architecture from Berkeley University PRISM (Power-efficient, Robust, hIgh-compression, Syndrome-based Multimedia coding)[4][5] in 2002 and finally European project DISCOVER (DIStributed Coding for Video SERvices)[6] based on Stanford architecture in 2007.

The Section 2 highlights implemented DVC codec Architecture and implementation details. The Section 3 presents the results with "C" code implementation, followed by conclusions and further work in Section 4.

# 2. DVC CODEC ARCHITECTURE AND IMPLEMENTATION DETAILS

The Architecture of implemented DVC codec is shown in Figure 1.

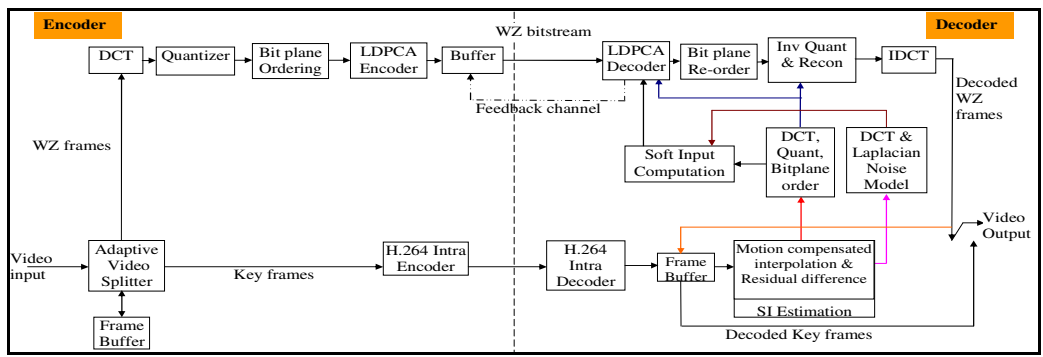

Figure 1. DVC Encoder and Decoder Architecture

## 2.1. DVC Encoder

The presented DVC encoder has following modules, which are explained in subsequent sub-sections:

1) Adaptive Video Splitter
2) Transform
3) Quantizer and Bit plane ordering
4) LDPCA Encoder and buffer
5) H.264 Intra Encoder

### 2.1.1. Adaptive Video Splitter

Adaptive Video Splitter is used to control the (non periodic) insertion rate of key frames in-between the WZ frames in an adaptive way. The GOP size control mechanism added to the encoder shall not significantly increase its complexity, i.e., shall not perform any motion estimation. The simple yet powerful metrics such as Difference of Histograms, Histogram of difference, Block Histogram difference and Block variance difference[7][8] are used to evaluate the motion activity along the video sequence.

### 2.1.2. Transform

The transform enables the codec to exploit the statistical dependencies within a frame, thus achieving better RD performance. It has been determined to deploy H.264 intra coder due to its lower bit rates, for key frames path. Hence the obvious choice of transform in WZ path is DCT to match with that of H.264. The WZ frames are transformed using a 4x4 DCT by breaking down image into 4x4 blocks of pixels, from left to right and top to bottom.

Once the DCT operation has been performed on all the 4x4 samples of image, the DCT coefficients are grouped together according to the standard Zig-Zag scan order[9] within the 4x4 DCT coefficient blocks. After performing the Zig-Zag scan order, coefficients are organized into 16 bands. First band containing low frequency information is often called the DC band and the remaining bands are called AC bands which contains the high frequency information.

### 2.1.3. Quantizer and Bit plane ordering

To encode WZ frames, each DCT band is quantized separately using predefined number of levels, depending on the target quality for the WZ frame. DCT coefficients representing lower spatial frequencies are quantized using uniform scalar quantizer with low step sizes, i.e., with higher number of levels. The higher frequency coefficients are more coarsely quantized, i.e., with fewer levels, without significant degradation in the visual quality of the decoded image. AC bands are quantized using dead Zone Quantizer with doubled zero interval to reduce the blocking artifacts. This is because the AC coefficients are mainly concentrated around the zero amplitude. The dynamic range of each AC band is calculated instead of using fixed value. This is to have quantization step size adjusted to the dynamic range of each band. The dynamic data range is calculated separately for each AC band to be quantized, and transmitted to the decoder along with the encoded bit stream. Depending on the target quality and data rates different types of quantization matrices given in Figure 2 can be used.

After quantizing the DCT coefficient bands, the quantized symbols are converted into bit-stream, the quantized symbols bits of the same significance (Ex: MSB) are grouped together forming the corresponding bit plane, which are separately encoded by LDPCA encoder.

Figure 2. Eight quantization matrices associated to different RD performances

### 2.1.4. LDPCA Encoder and buffer

It is determined to employ Low Density Parity Check Accumulator (LDPCA)[10] channel encoder (aka WZ encoder), as it performs better with lower complexity compared to turbo codes[10]. An LDPCA encoder consists of an LDPC syndrome-former concatenated with an accumulator. For each bit plane, syndrome bits are created using the LDPC code and accumulated modulo 2 to produce the accumulated syndrome. The encoder stores these accumulated syndromes in a buffer and initially transmits only a few of them in chunks. If the decoder fails, more accumulated syndromes are requested from the encoder buffer using a feedback channel. To aid the decoder detecting residual errors, an 8-bit CRC sum of the encoded bit plane is also transmitted.

### 2.1.5. H.264 Intra Encoder

Key frames are encoded by using H.264 Intra (Main profile)[10] coder. Coding with H.264/AVC in Main profile without exploiting temporal redundancy is intra coding. And H.264 intra coding is among the most efficient intra coding standard solutions available, even better than JPEG2000 in many cases. The JM reference software[11] is used as main profile intra encoder in this implementation. Quantization Parameters (QP) for each RD point are chosen to match with the WZ frame quality.

### 2.2. DVC Decoder

The presented DVC decoder has following modules, which are explained in subsequent sub-sections:

1) H.264 Intra Decoder

2) SI Estimation

3) DCT and Laplacian noise model

4) Soft Input Computation

5) LDPCA Decoder

6) Inverse Quantizer and Reconstruction

7) IDCT

### 2.2.1. H.264 Intra Decoder

Similar to the H.264 Intra encoder, decoder (Main profile) specifications and reference software are used from [9] and [11] respectively. These key frames decoded outputs are used to estimate the Side information.

### 2.2.2. SI estimation

A frame interpolation algorithm[7] is used at the decoder to generate the side information. When correlation between side information and WZ frame to be encoded is high, fewer parity bits needs to be requested from encoder to reach a certain quality level. Another important issue to be considered in the motion interpolation framework is capability to work with longer GOP without significant degradation in quality of interpolated frame, especially when correlation of frames in the GOP is high. This is a complex task since interpolation and quantization errors are propagated inside the GOP, when frame interpolation algorithm uses WZ decoded frames as reference. This frame interpolation framework works for GOP of any length including longer and high motion GOP. Figure 3 shows details of frame interpolation scheme.

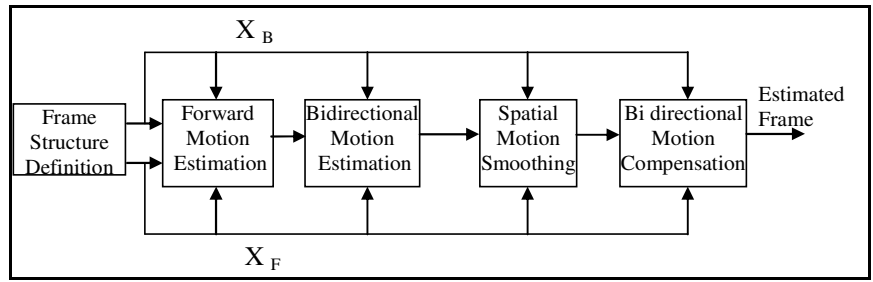

Figure 3.  Side information estimation

The frame interpolation structure used to generate the side information is based on previously decoded frames, $X_B$ and $X_F$, the backward (in the past) and forward references (in the future). An example frame structure definition for GOP=4 is shown in Figure 4.

In forward motion estimation, a full search based block motion estimation algorithm is used to estimate the motion between the decoded frames $X_B$ and $X_F$. The search for the candidate block is exhaustively carried out in all possible positions within the specified range ±M (M=32) in the reference frame.

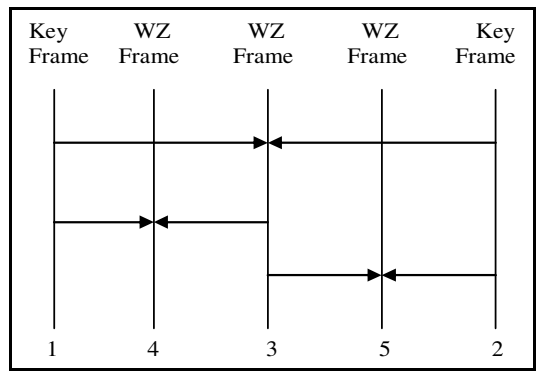

Figure 4.  Frame structure definition

The bidirectional motion estimation[7][12] module refines the motion vectors obtained in the forward motion estimation step by using a bidirectional motion estimation scheme similar to the "B" frame coding mode used in current video standards[13]. But here as the interpolated pixels are not known, a different motion estimation technique is used. This technique selects the linear trajectory between the next and previous key frames passing at the centre of the blocks in the interpolated frame as shown in Figure 5.

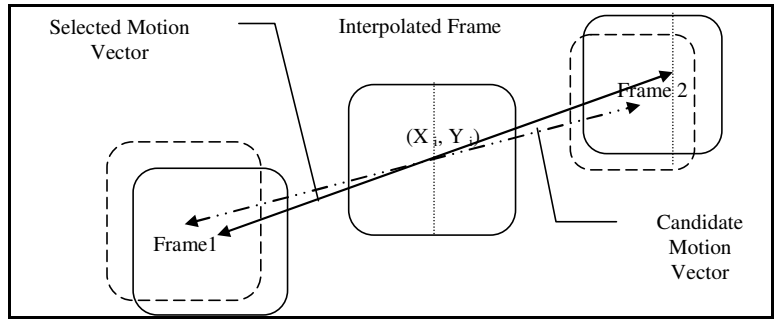

Figure 5.  Bidirectional Motion Estimation

This bidirectional motion estimation technique combines a hierarchical block size (first 16x16 and then followed by 8x8), with an adaptive search range in the backward and forward reference frames based on the motion vectors of neighbouring blocks.

Extensive experimentation has been carried out to find out whether half-pel for forward as well as bidirectional motion estimation works well or any other combination with integer-pel. Here, we propose to use integer-pel for forward motion estimation and half-pel motion estimation for 16x16 and 8x8 bidirectional motion estimation, which gave best results.

Next, a spatial motion smoothing algorithm[14] is used to make the final motion vector field smoother, except at object boundaries and uncovered regions by using weighted vector median filters, and by evaluating prediction error and the spatial properties of the motion field.

Once the final motion vector field is obtained, the interpolated frame can be filled by simply using bidirectional motion compensation as defined in standard video coding schemes[13].

### 2.2.3. DCT and Laplacian noise model

In DVC, decoding efficiency of WZ frame critically depends on the capability to model the statistical dependency[15][16] between the WZ information at the encoder and the side information computed at the decoder. This is a complex task since the original information is not available at the decoder and the side information quality varies throughout the sequence, i.e., the error distribution is not temporally constant. A Laplacian distribution, which has good tradeoff between model accuracy and complexity, is used to model the correlation noise, i.e., the error distribution between corresponding DCT bands of SI and WZ frames. The Laplacian distribution parameter is estimated online at the decoder which takes into consideration the temporal and spatial variability of the correlation noise statistics. The technique used estimates the Laplacian distribution parameter "$\alpha$" at the DCT band level (one "$\alpha$" per DCT band) and at the coefficient level (one "$\alpha$" per DCT coefficient). The estimation approach uses the residual frame, i.e., the difference between $X_B$ and $X_F$ (using the motion vectors), as a confidence measure of the frame interpolation operation, and also a rough estimate of the side information quality.

### 2.2.4. Soft input computation

The conditional bit probabilities are calculated by considering the previously decoded bit planes and the value of side information (SI). The benefits of LDPCA codes are that they incorporate the underlying statistics of the channel noise into the decoding process in the form of soft inputs or priori probabilities. The probability calculations are different for DC (top left corner low frequency coefficient of each 4x4 DCT block) and AC (remaining high frequency coefficients of each 4x4 DCT block) bands because DC band contains only unsigned coefficients, whereas AC bands contain signed coefficients.

DC band bit probability calculations (of bit being either 0 or 1) depends on previously decoded bit planes and SI DCT coefficients as given below:

a) DC band bit probability calculation for current bit plane $b^{th}$ bit being '0':

$$P\!\left(Q^{(b)} = 0 \mid Y = y, Q^{(b+1)}, \ldots\ldots Q^{(L-1)}\right) = \sum_{i = x_p * W}^{(x_p + 2^b) * W - 1} \frac{\alpha}{2} e^{-\alpha |i - y|} \qquad (1)$$

b) DC band probability calculation for current bit plane $b^{th}$ bit being '1':

$$P\!\left(Q^{(b)} = 1 \mid Y = y, Q^{(b+1)}, \ldots\ldots Q^{(L-1)}\right) = \sum_{i = (x_p + 2^b) * W}^{(x_p + 2^b + 2^b) * W - 1} \frac{\alpha}{2} e^{-\alpha |i - y|} \qquad (2)$$

Where,
$$x_p = \sum_{i = b+1}^{L-1} Q^{(i)} . 2^i \qquad (3)$$

Further, "y" is SI DCT coefficient for current bit, "W" is the quantizer step size and "b" is the current bit plane to be coded, "L" is the total number of bits reserved for current coefficient.

For AC bands, extensive experimentation has been carried out to determine which is best among SI quantization or SI DCT coefficients, for probability calculations. For AC bands, calculation of the probability (for input bit-plane bit being 0) is calculated by considering the information of previously decoded bit planes and SI DCT.

$$P\!\left(Q^{(b)} = 0 \mid Y = y, Q^{(b+1)}, \ldots\ldots Q^{(L-1)}\right) = \sum_{i = x_p * W}^{(x_p + 2^b) * W - 1} \frac{\alpha}{2} e^{-\alpha |i * sign - y|} \qquad (4)$$

If $(L-1)^{th}$ bit is of bit plane is "1" then "sign" becomes "-1" otherwise "sign" is "1".

Here, we propose that calculation of probability (for input bit-plane bit being 1) shall be done differently for MSB (represents sign) and other bits. MSB bit probability (of bit being 1) calculations shall be done using previously decoded bit planes and SI quantized values as given below:

$$P\!\left(Q^{(L-1)} = 1 \mid Y = y\right) = \sum_{i=1}^{2^b - 1} \frac{\alpha}{2} e^{-\alpha |-i - y_q|} \qquad (5)$$

Here, $y_q$ is the quantized side information.

And all other bits probability (of bit being 1) calculations shall be carried out using previously decoded bit planes and SI DCT coefficients (which gave best RD performance), as shown below:

$$P\left(Q^{(b)} = 1 \mid Y = y, Q^{(b+1)}, \ldots\ldots Q^{(L-1)}\right) = \sum_{i=x_p*W}^{(x_p+2^b)*W-1} \frac{\alpha}{2} e^{-\alpha|i*sign-y|} \qquad (6)$$

Using these probabilities a parameter called Log likelihood Ratio (LLR) intrinsic is calculated. Using LLR intrinsic value LDPCA decoder decodes the current bit plane.

### 2.2.5. LDPCA decoder

This decoding procedure[10][17] explains decoding of a bit plane given the soft input values of side information and the parity bits transmitted from the encoder. This procedure is repeated for every increment in the number of parity bit requests from the decoder. Before the decoding procedure starts the syndrome bits are extracted from received parity bits from encoder, by doing inverse accumulation operation according to the graph structure[18]. Sum-product decoding operation is performed on these syndrome bits. This algorithm is a soft decision algorithm which accepts the probability of each received bit as input. To establish if decoding is successful, syndrome check error is computed, i.e. the Hamming distance between the received syndrome and the one generated using the decoded bit plane, followed by a cyclic redundancy check (CRC). If the Hamming distance is non-zero, then the decoder proceeds to the next iteration and requests more accumulated syndromes via the feedback channel. If the Hamming distance is zero, then the successfulness of the decoding operation is verified using the 8-bit CRC sum. If the CRC sum computed on the decoded bit plane matches to the value received from the encoder, the decoding is declared successful and the decoded bit plane is sent to the inverse quantization & reconstruction module.

### 2.2.6. Inverse Quantizer and Reconstruction

Inverse quantization is carried out after the successful decoding of the all bit planes of a particular band. For each coefficient of the bit planes are grouped together to form quantized symbols. After forming the quantization bin, each bin tells the decoder where the original bin lies, i.e. in an interval. The decoded quantization bin is an approximation of the true quantization bin obtained at the encoder before bit plane extraction.

Here, we propose to use different types of reconstruction methods for positive, negative and zero coefficients, as described below:

a)   If decoded bin q>0:

Inverse quantized coefficient range is calculated using $Z_i=q.W$ and $Z_i+1= (q+1).W$, where "q" is decoded bin and "W" is the step size. The reconstructed coefficient is calculated using below equation, where "y" is SI DCT coefficient:

$$X_{re} = \begin{cases} Z_i, & y < Z_i \\ y, & y \in [Z_i, Z_{i+1}) \\ Z_{i+1} & y \geq Z_{i+1} \end{cases} \qquad (7)$$

b)   If decoded bin q<0:

For this bin, decoded quantized bin range is calculated using $Z_i=q.W$ and $Z_i+1= (q-1).W$. The reconstructed coefficient is calculated using below equation:

$$X_{re} = \begin{cases} Z_i, & y \rangle Z_i \\ y, & y \in [Z_i, Z_{i+1}) \\ Z_{i+1} & y \leq Z_{i+1} \end{cases} \qquad (8)$$

158

c)   If decoded bin q=0:

For this bin, decoded quantized bin range is calculated using $Z_i$=-W and $Z_i$+1= W. The reconstructed coefficient is calculated using below equation:

$$X_{re} = \begin{cases} Z_i, & y \leq Z_i \\ y, & y \in (Z_i, Z_{i+1}] \\ Z_{i+1} & y \rangle Z_{i+1} \end{cases} \qquad (9)$$

### 2.2.7. Inverse DCT

IDCT operation is carried out after performing inverse quantization and inverse zig-zag scan. IDCT operation is to restore the image into pixel domain from transform domain.

## 3. DVC "C" MODEL IMPLEMENTATION RESULTS

The DVC encoder and decoder described are completely implemented in "C". The implemented codec has been evaluated with four standard test sequences namely QCIF hall monitor, coast guard, foreman and soccer sequences with 15 Hz vertical frequency. The chosen test sequences are representative of various levels of motion activity. The Hall monitor video surveillance sequence has low to medium amount of motion activity. And Coast guard sequence has medium to high amount of motion activity, whereas Foreman video conferencing sequence has very high amount of motion activity. And Soccer sequence has significant motion activity. The H.264 coder profile in key frame path is main profile, which can encode only 4:2:0 sequence, and not 4:0:0. All the metrics were measured only for the luma component of video sequences. Hence chroma components of test sequences are replaced with 0's and used H.264 in 4:2:0 mode for luma analysis. Figures 6 to 9 shows RD performance and comparison with that of H.264 AVC (Intra) and H.264 AVC (No motion), with fixed GOP of 2.  The "No motion" mode exploits temporal redundancy in IB…IB… structure but without performing motion estimation, which is the most computationally intensive encoding task.

For Hall monitor sequence, PSNR achieved by DVC is around 2-3dB better than that of H.264 intra and upto 2dB lower than that of H.264 "no motion" for a given bit rate as shown in Figure 6. For Coast Guard sequence, PSNR achieved by DVC is around 1-2dB better than that of H.264 intra and around 0.5-1.8dB better than that of H.264 "no motion" for a given bit rate as shown in Figure 7. The RD performance achieved for Q1-Q3 quantization levels is better than that of shown in reference [20] by 16kbps and 0.3dB.

For Foreman sequence, PSNR achieved by DVC is around (-1) → (+1) dB lower/better than that of H.264 intra and around 0.5-2.2dB lower than that of H.264 "no motion" for a given bit rate as shown in Figure 8. Further, DVC PSNR is better than that of H.264 intra for lower quantization parameters, whereas for higher quantization parameters it is not. This is due to lower SI quality with higher quantization parameters with this sequence. The RD performance achieved for Q1-Q4 quantization levels is better than that of shown in reference [20] by 20kbps.

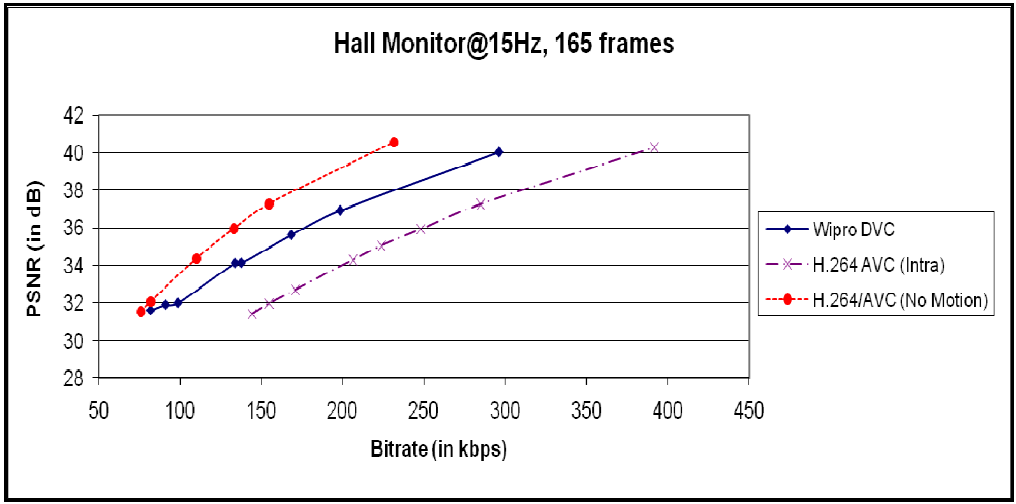

Figure 6.  Hall Monitor RD performance

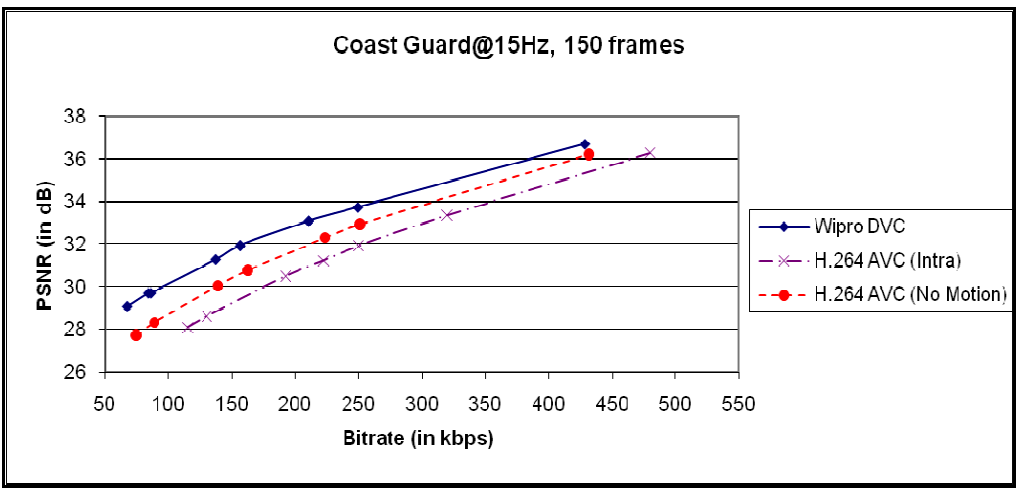

Figure 7.  Coast Guard RD performance

For Soccer sequence, PSNR achieved by DVC is around 3dB lower than that of H.264 intra and H.264 "no motion" for a given bit rate as shown in Figure 9. The DVC RD performance is lower with this sequence, due to significant motion activity which causes lower SI quality.

Also as expected, H.264 "no motion" RD performance is better than that of H.264 intra with all the video sequences. And the PSNR gain with H.264 "no motion" compared to H.264 intra is a function of motion activity, i.e., with increasing motion activity PSNR gain reduces. For example, significantly high motion activity soccer sequence RD performance of H.264 "no motion" and H.264 intra are comparable.

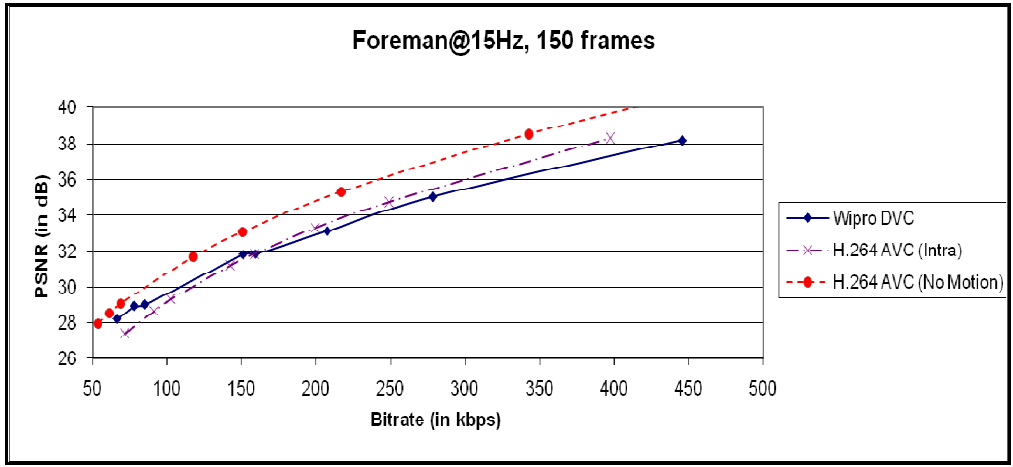

Figure 8.  Foreman RD performance

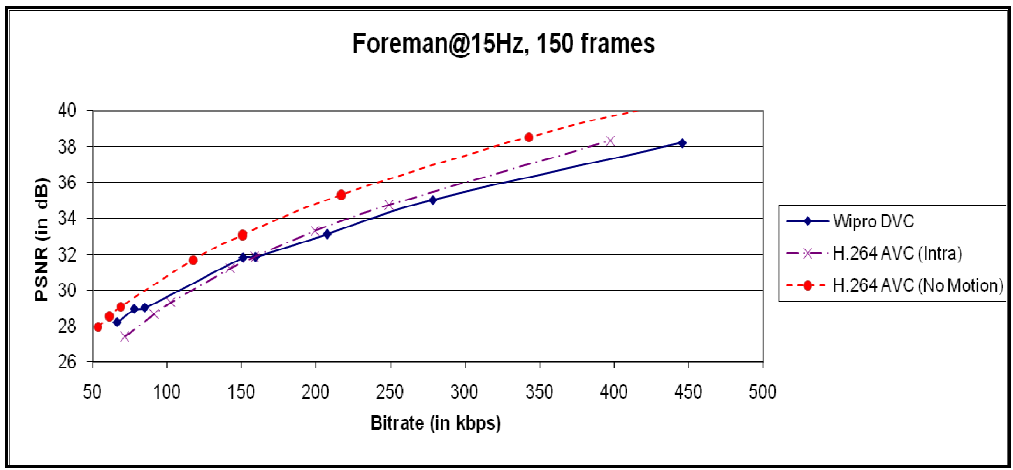

Figure 9.  Soccer RD performance

# 4. CONSLUSION & FUTURE WORK

The DVC Architecture, implementation details and results are presented in this paper. It has been proven that DVC resulted in better RD performance than that of H.264 intra for low to medium motion activity sequences. For high motion sequences, the RD performance of DVC and H.264 intra are comparable. Whereas, DVC RD performance of very high motion sequences are lower than that of H.264 intra. We have highlighted gaps and challenges of DVC in practical usage, in another paper [1]. In summary, following are the key gaps and challenges in practical usage of DVC in the present state:

1) **Feedback channel from decoder to encoder:** In some applications such as scanner-on-the-go, there may not be any physical feedback channel available. Even if a feedback channel is available, the decoding delay will be more due to iterative decoding process and only online decoding is possible. Hence it is highly desired to eliminate feedback channel by using rate estimation at the encoder

2) **Lack of procedure for chroma components coding:** In literature, there is no mention of encoding and decoding procedure for chroma component. And even performance comparison has been done only for Luma component

3) **Lack of compressed video transport bit-stream definition:** In literature WZ bit stream and H.264 bit stream are shown sent separately to decoder. But for practical applications, we need a combined transport bit-stream definition for combined WZ and H.264

4) **Inconsistencies in RD performance compared to H.264 intra with different video streams:** DVC is performing better than H.264 intra only upto certain motion activity video sequences. And for significantly high motion activity video sequences, DVC performance is poorer than that of H.264 Intra. In the light of this, it is desirable to detect significantly high motion frames and use H.264 intra only coding for such frames

5) **Flicker:** H.264 intra decoded frames are de-blocking filtered, where-as WZ decoded frames are not. This causes flicker when H.264 and WZ decoded frames are shown in interleaved way based on GOP

6) **No standardization yet:** As there is no standardization yet for DVC, there are differing implementations. For the sake of interoperability standardization is a must before DVC sees practical usage in a broader way

Our further work includes working on the gaps and challenges highlighted in [1] and include those schemes worked out into already implemented codec presented in this paper. Also DVC presented in this paper is mono-view DVC. And the methods can be easily extended to multi-view DVC.

**Authors**

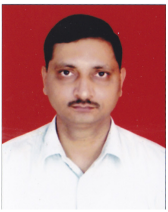

**Vijay Kumar Kodavalla** is presently working as Lead Architect in Wipro Technologies, Bangalore, India. He has more than 16 years of VLSI industry experience, worked on complex ASIC, SoC, FPGA and System designs in Audio/Video domain. He has three US Patents pending and presented several papers in international conferences. He has obtained B.Tech (Electronics & Comm.), M.Tech (Electronics & Comm.) and MS (Micro Electronics) degrees in 1992, 1995 and 2002 years respectively. He is currently pursuing PhD from JNTU College of Engineering, Hyderabad. His research interests include Image/Video compression, post processing Distributed Video Coding and Chip Architectures.

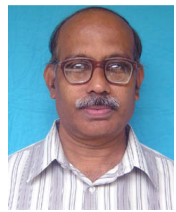

**Dr. P.G. Krishna Mohan** is presently working as Professor in JNTU College of Engineering, Hyderabad, India. He has more than 29 years of teaching experience, out of which 15 years of research experience. He has 21 publications in various Journals & Conferences at national and international level. Two person obtained Ph.D under his guidance and presently guiding six students for Ph.D and one student for M.S by research. He has obtained B.Tech. (ECE) R.E.C Warangal, M.E (Comm. System) Roorkee University, Ph.D (Signal Processing) IISc Bangalore. His research interests include Analysis & Design in Signal Processing and Communication Engineering.